\documentclass[epjCONF]{svjour}
\usepackage{graphicx}
\usepackage[varg]{txfonts} % Times fonts
\usepackage[latin1]{inputenc}
\usepackage{subfig}
\session-title{UHECR 2012, CERN, Feb. 13-17, 2012}
\begin{document}
\title{A new physical phenomenon in \\ ultra-high energy collisions}
\author{Glennys R. Farrar\inst{1}\fnmsep\thanks{\email{gf25@nyu.edu}} \and Jeffrey D. Allen\inst{1}  }
\institute{Center for Cosmology and Particle Physics \&
Department of Physics, New York University, New York, NY 10003, USA}
\abstract{
%Ultrahigh energy cosmic rays (UHECRs) are the highest energy particles in the Universe, recorded to have energies up to 100 million times greater than the LHC beams, resulting in CR-air collisions with center-of-mass energies up to 3000 TeV.  %Whether the UHECRs are protons or nuclei and how they are accelerated are presently uncertain.   
We show that combining the published Pierre Auger Observatory measurements of the longitudinal \emph{and} lateral properties of UHE atmospheric showers, points to an unforeseen change in the nature of particle interactions at ultrahigh energy.  A ``toy model" of UHE proton-air interactions is presented which provides the first fully consistent description of air shower observations.  It demonstrates that the observed energy dependence of the depth-of-shower-maximum distribution may not indicate a transition to a heavier composition, as commonly assumed. While fundamentally phenomenological, the model is based on considerations of how the normal vacuum of QCD might be vaporized and chiral symmetry restored by the extreme energy densities produced in UHE collisions.  Whatever its origin, understanding this unexpected phenomenon opens exciting directions in particle physics and may impact Early Universe cosmology. 
} %end of abstract
\maketitle
\section{Preliminaries}
\label{intro}
Thousands of cosmic ray air showers with energies above a few EeV (1 EeV = $10^{18}$ eV) have been observed by the Pierre Auger Observatory simultaneously in the Fluorescence and Surface Detectors (FD \& SD) \cite{augerNIM04,augerEnergyCalibPLB10}.  The fluorescence technique allows the longitudinal profile of the shower to be measured, while the surface detector samples the lateral distribution function of produced particles.  The FD signal is predominantly due to electromagnetic particles ($e^\pm$ and photons) which are concentrated in the core of the shower and carry $\sim 90\%$ of the total energy, while the Surface Detector samples shower particles at distances typically a kilometer and more from the core.  The strength of the ground signal is characterized by $S(1000)$, its value at 1000 m;  at large radii and especially at large zenith angles, the SD signal is mostly due to muons from pion decay.  The Auger analysis is now very mature and well-understood, so that the combined information provided by surface array and fluorescence telescope observations give an extremely powerful constraint on the primary CRs and their interactions.%; for instance excluding that more than a small fraction of UHECRs are photons\cite{augerPhotons&ExoticsICRC11}.  

It has been known for some time that existing UHECR shower simulations fall short of accounting for the muon content of the ground signal, by a significant factor \cite{Engel_HadInt_ICRC07,jaICRC11}.   We show here that -- no matter what the composition of the UHECR primaries -- any presently available model based on conventional particle physics cannot simultaneously explain the totality of Auger observations on the longitudinal development and lateral shape of the showers.  A model that reproduces the distribution of where in the atmosphere the showers peak, cannot explain the magnitude {\em and} zenith angle dependence of the ground signal, unless the fraction of energy carried by the hadronic shower is approximately a factor-two greater than in conventional models;  as far as we can see, this requires a strong suppression in decaying $\pi^0$'s at ultrahigh energies compared to predictions of standard physics.  

% In order to reconcile theory and observations, either the production or decay of $\pi^0$'s must be suppressed dramatically at UHE.  We propose a simple phenomenological model which accounts in remarkable detail for all of the observed shower properties.  The primary cosmic rays are taken to be protons, and pion production is suppressed as a result of the restoration of chiral symmetry in ultrahigh energy collisions.   We discuss the difficulties with most other ``new physics" solutions but note that Lorentz Invariance Violation or other mechanism to suppress $\pi^0$ decay may provide another approach to an explanation. 

\begin{figure}[t]
\centering
\includegraphics[width=0.6\textwidth]{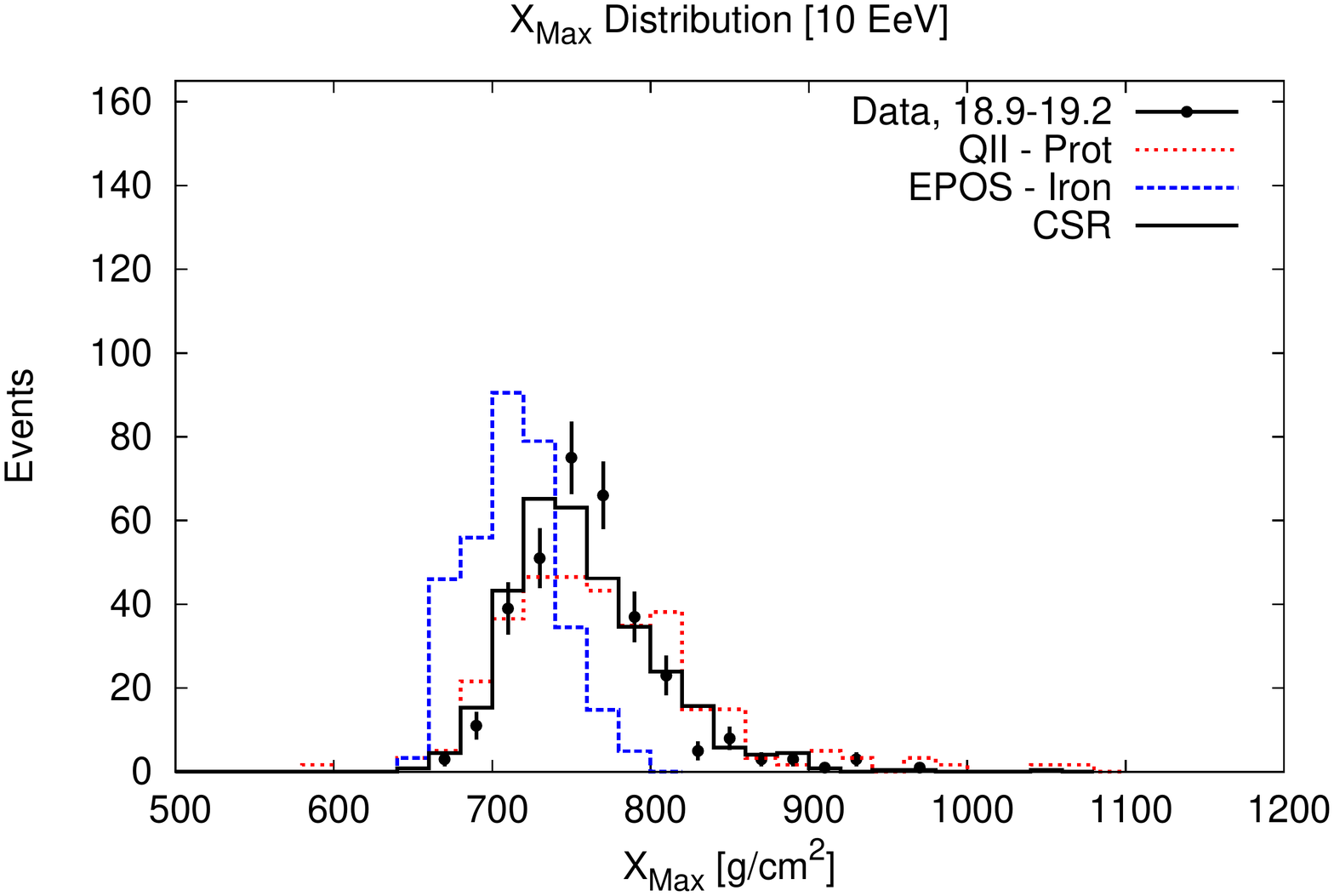}
\includegraphics[width=0.7\textwidth]{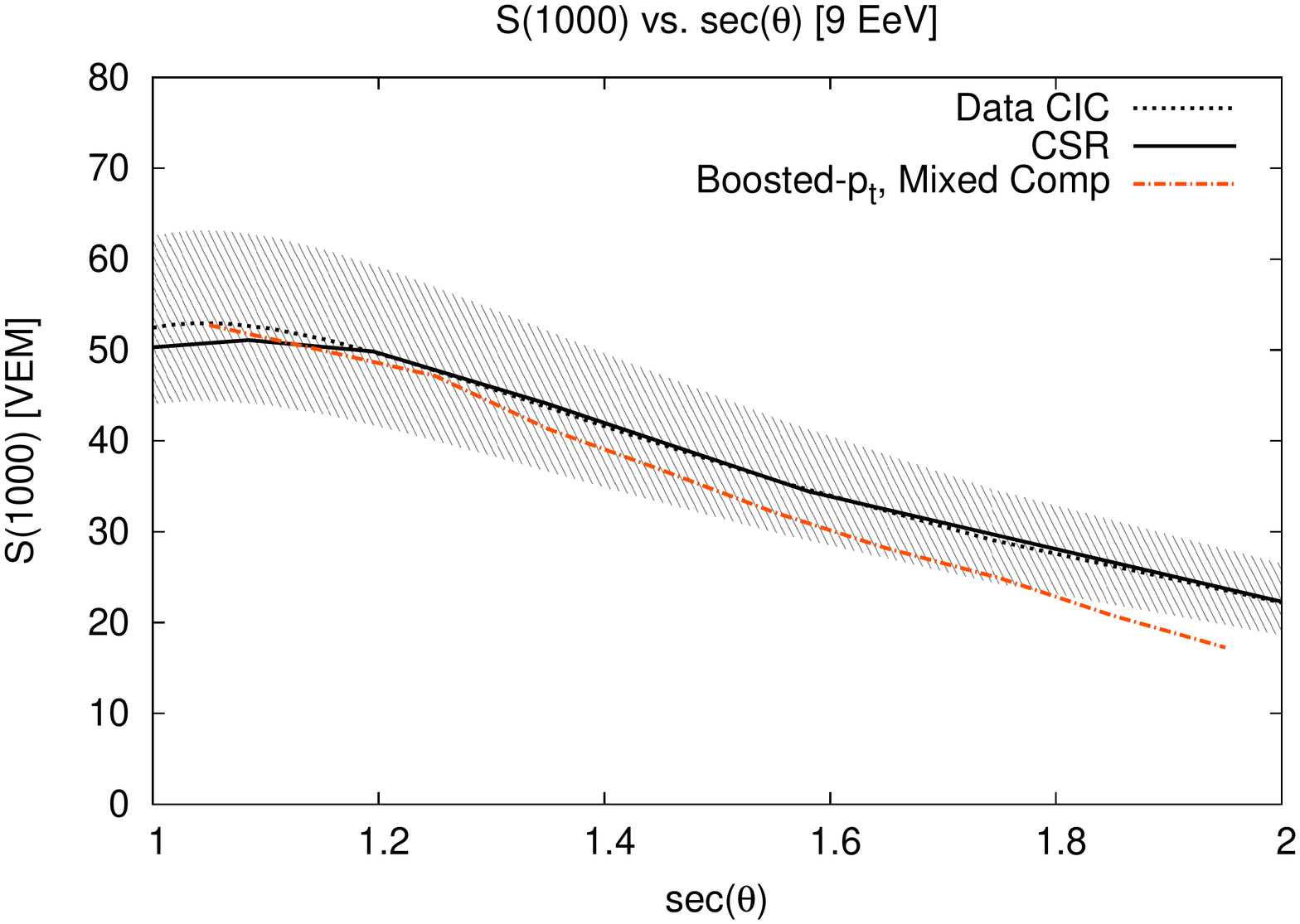}
\caption{(top): The measured distribution of $X_{\rm max}$ in the energy range $10^{18.9-19.2}$ eV from Ref.\ \cite{augerXmax,augerXMaxPRL}, and predictions (smeared with the $\pm 20 {\rm g cm}^{-2}$ FD measurement resolution) of the standard event generators EPOS with Fe (blue-dashed) and QGSJet II with protons (red-dotted), and the new CSR model developed here (black-solid). (rhs): The measured ground signal as a function of zenith angle at $E_{FD} = 9$ EeV (black dotted line), calculated using the equations and best fit parameter values from Ref.\ \cite{augerCIC_ICRC11}; the grey band represents the impact of the 22\% systematic uncertainty in the energy scale.  Also shown are predictions from the Chiral Symmetry Restoration (black solid) and Mixed Composition-with-boosted-$p_t$ (orange dash-dot) models developed here.  Because the relationship between the calorimetric energy in the FD and the true energy is slightly different in the CSR model than in the models used by Auger to determine the missing energy, the corresponding true energy scale is approximately 10 EeV in the CSR model.  The predicted CIC curve -- ground signal as a function of zenith angle -- at $E_{FD} = 9$ EeV is shown for, top to bottom, the Chiral Symmetry Restoration CSR model (black solid), Mixed Composition-with-boosted-$p_t$ (orange heavy dash-dot),  EPOS-Fe (blue dash),  EPOS-Mixed (grey light dash-dot), and QII-p (red short-dash) models.  \label{Xmax-CIC}}
\end{figure}

The depth in the atmosphere of the first interaction is controlled by the UHECR-air nucleus cross-section and typically occurs for protons when the cosmic ray has penetrated a column depth of $\approx {40\, \rm g \, cm}^{-2}$ -- about 20 km above ground level for vertical showers.  Hundreds to thousands of secondary particles are usually produced, many of which also have energies above the highest accelerator energies.  These secondaries are essentially all hadrons (particles composed of quarks and/or anti-quarks); based on final states observed in accelerator experiments, these secondaries are expected to mainly be pions.  With a lifetime of $8.4 \times 10^{-17}$\,s, $\pi^0$'s decay immediately to two photons, feeding the electromagnetic component of the shower, while the charged pions and other hadrons propagate until they interact or decay.   Relativistic time dilation increases the decay length by the factor $E/m c^2$, so high energy particles other than $\pi^{0}$'s generally interact rather than decaying, continuing the hadronic shower, while below $\approx$100 GeV most $\pi^\pm$'s decay before interacting.  At each successive step in the cascade, energy is removed from the hadronic shower and injected into the electromagnetic (EM) shower via the decays of produced $\pi^0$'s.   The integrated longitudinal profile (LP) provides a calorimetric measurement of the energy of the primary CR, with a relatively small uncertainty due to the correction for energy lost to neutrinos and particles hitting the ground.

The peak in the longitudinal profile observed by the FD occurs at a column depth designated $X_{\rm max}$, typically 700-800 g cm$^{-2}$, where the number of $e^\pm$ in the EM shower is at a maximum.  The maximum muon production in the shower occurs at $X_{\mu, \, \rm max}\approx 550 \,{\rm g \,cm}^{-2}$, and the mean depth of production of pions decaying to muons detected in the SD at 1 km is $X_{\pi-\mu} \approx 600 \,{\rm g \,cm}^{-2}$ -- about 5 km altitude for vertical showers.  The value of $X_{\rm max}$ depends on the depth of first interaction and the rate at which the shower develops.  The latter depends on the elasticity (fraction of incoming energy carried by the leading secondary particle) and the multiplicity of secondary particles in the early, high energy interactions.  These vary a lot from event to event for proton primaries, so the $X_{\rm max}$ distribution for proton primaries is quite broad.  The $X_{\rm max}$ distribution of nuclei is both shallower and narrower than for protons, because the nuclear cross section $\sim A^{2/3}$ and the final state multiplicity is higher.  For illustration, Fig.\ \ref{Xmax-CIC}(top) shows the distribution of $X_{\rm max}$ for events in the energy range $10^{18.9 - 19.2}$ eV from combining the two adjacent bins in Fig.\ 3 of Ref.\ \cite{augerXmax}, along with representative predictions of the most popular event generators used in air shower simulations at present, QGSJet II-3 \cite{QIIv3} for protons (QII-p) and EPOS 1.99 \cite{EPOS} for iron (EPOS-Fe), and the distribution from the Chiral Symmetry Restoration (CSR) toy model discussed below.

High energy interactions dictate the energy partition between the electromagnetic and hadronic components of any given atmospheric shower, with the energy in the hadronic shower governing the number of muons at ground.   The number of muons at ground is roughly speaking just the energy in the hadronic shower divided by 100 GeV, since most secondaries are pions and almost all charged pions below 100 GeV decay and produce a muon before reaching ground.  This is a simplified picture to aid intuition, and of course actual shower simulations take into account the baryon and kaon content of the shower, non-decaying pions, muon attenuation, etc.  This final stage of the shower development is accurately and reliably modeled, as it involves physics that is well-understood from accelerator measurements.

There are two possible ways to increase the muon content predicted by simulations, to improve agreement with the Auger SD observations:
\begin{enumerate}
\item Increase the total number of muons by increasing the fraction of energy in the hadronic component of the shower \cite{grieder73,pierogWerner08}.  
\item Increase the production angle of muons so they travel further from the core before reaching ground;  the muon lateral distribution function (LDF) falls with radius, so this would increase the muon number at 1 km.  
\end{enumerate}
Because the distribution of final state particles in $\pi^\pm$-air collisions is not yet available in the entire range of incident energies and phase space which may be relevant, the latter option may be possible without incompatibility with experimental data;  the NA61/SHINE experiment at CERN is working to fill this gap. 

We have performed simulations modifying the angular distribution of produced pions and hence muons, for a mixed composition which fits the $X_{\rm max}$ distribution at 10 EeV and minimizes the modification required in the event generators; we call this the ``boosted-$p_t$" model and will report details elsewhere.  Fitting $S(1000)$ for vertical showers requires a radical increase in the $p_t$ distribution of pions starting just above the highest energy in which the event generators are constrained by experiments.   Continuity with respect to measured final state distributions in $p$-C at 31 and 158 GeV incident momenta \cite{NA61pC31,NA49pC158} makes this scenario extremely implausible theoretically for vertical showers, and the required modification only increases for larger zenith angles.   Trying to modify the $p_t$ distribution is a moot exercise, however, because flattening the muon lateral distribution function can be ruled out directly from the observed zenith angle dependence of the ground signal.  
 
The so-called Constant Intensity Cut or CIC curve gives $S(1000)$ as a function of zenith angle, for any specified range of UHECR energy.  The shape of the CIC curve is particularly robust, because it is completely model independent.  It relies only on the fact that the flux of CRs at any given energy is independent of zenith angle and that $S(1000)$ is monotonic with energy, apart from measurement variance.  Thus the top $N$ events in bins of constant solid angle have the same energies independent of zenith angle, as long as $N$ is large enough that sample variance is not an issue.   To obtain the normalized CIC curve for a given primary UHECR energy bin requires an absolute energy calibration which Auger achieves using the FD energy in hybrid events \cite{augerCIC_ICRC11}.   The resulting CIC curve has an overall calibration uncertainty of $\approx 20$\%, dominated by the FD systematic uncertainties.  But thanks to the angular resolution of better than 1 degree \cite{augerAngRes} and the high statistics in the SD sample at 10 EeV and below, the shape of the CIC curve is well-determined.  The shape of the CIC curve reflects how the shape of the LDF changes with zenith angle.  Fig.\ \ref{Xmax-CIC}(rhs) shows the CIC curve measured by Auger at $E_{FD} = 9$ EeV from \cite{augerCIC_ICRC11}, along with predictions of the boosted-$p_t$ and CSR models to be explained below;  throughout, we simulate the detector response and event reconstruction using the Auger Offline analysis software package \cite{offline07}.  The grey hatched error band indicates the absolute calibration uncertainty but there is little uncertainty in the shape.

\begin{figure}[t]
\centering
\includegraphics[width=0.49\textwidth]{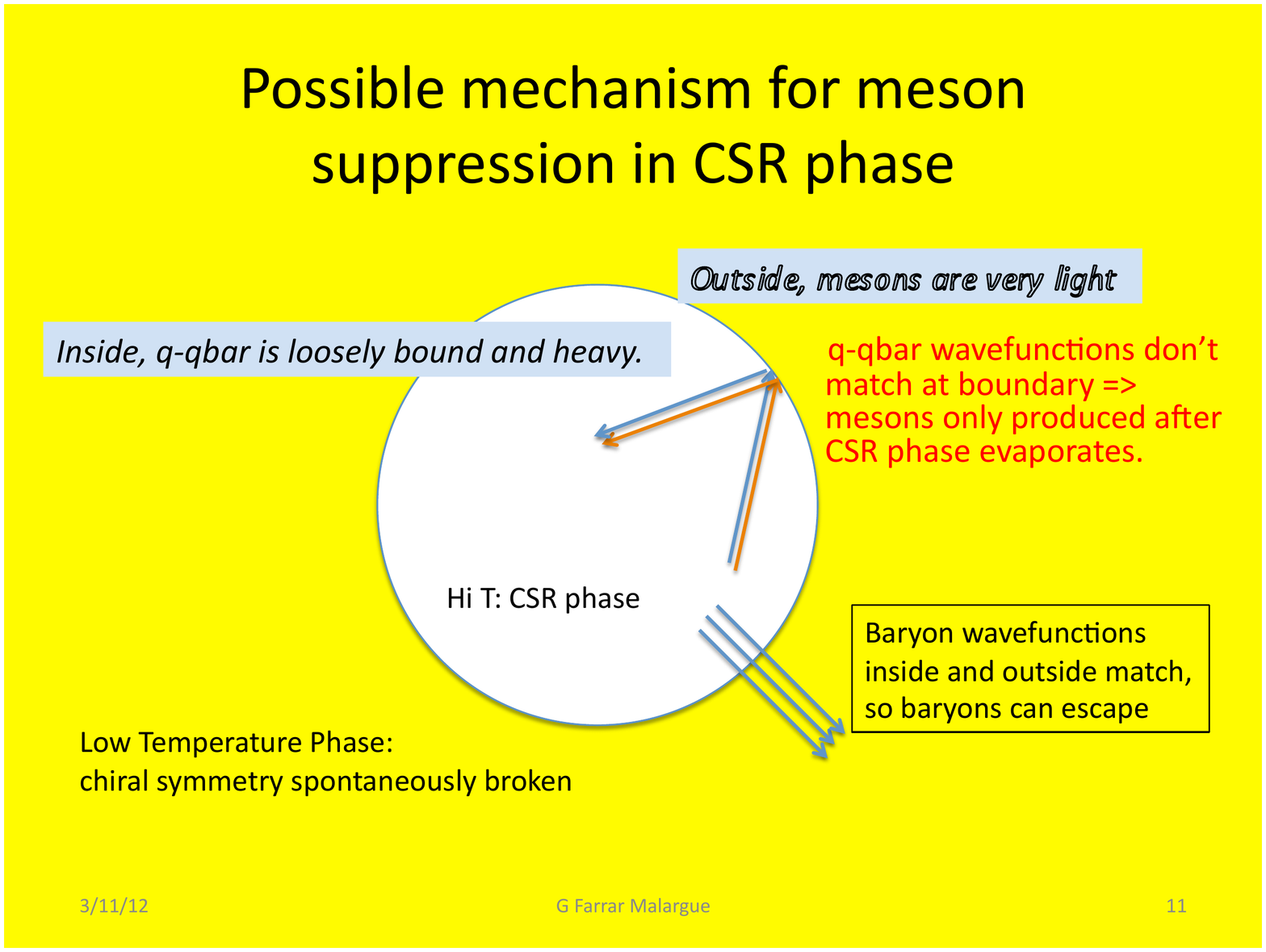}
\includegraphics[width=0.49\textwidth]{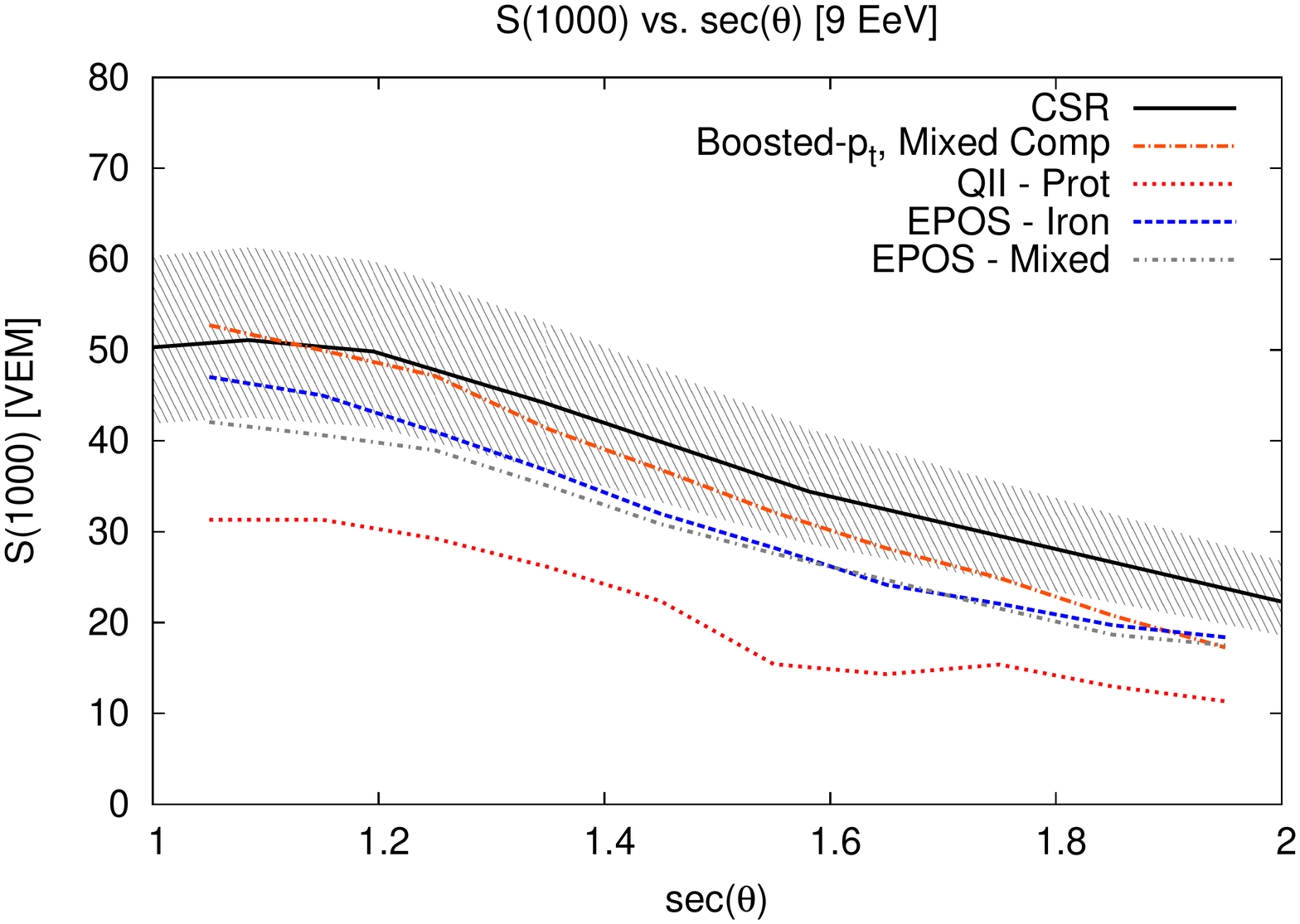}
\caption{ (lhs): Illustration of a mechanism by which meson production may be suppressed.  If the wave function of a $q \bar{q}$ bound state in the CSR phase has little overlap with a meson in the low temperature exterior phase, it will experience near-total internal reflection making the $q$ and $\bar{q}$ available for re-assembly into other bound states.  Oh the other hand, a $qqq$ or $\bar{q} \bar{q} \bar{q}$ bound state can be expected to have a larger overlap with the wave function of its corresponding baryon or anti-baryon and thus escape.  The color hyperfine interactions which are responsible for the mass splittings within hadron multiplets act to induce formation of bound states even though confinement is not operative. (rhs): The predicted CIC curve -- ground signal as a function of zenith angle -- at $E_{FD} = 9$ EeV for, top to bottom, the Chiral Symmetry Restoration CSR model (black solid), Mixed Composition-with-boosted-$p_t$ (orange heavy dash-dot),  EPOS-Fe (blue dash),  EPOS-Mixed (grey light dash-dot), and QII-p (red short-dash) models.  \label{CSR}}
\end{figure}

If the ground signal at 1 km is increased compared to a reference model, by flattening the muon LDF without increasing the total number of muons, the relative increase is less at larger zenith angles than for vertical showers.   The physical reason is that muons outside the core of the shower come from the collinear decay of pions which are produced at approximately the same characteristic depth, $X_{\pi-\mu}$, no matter what the zenith angle of the shower.  That is, the muons effectively radiate from a fixed depth on the shower axis, so muons arriving at 1 km radius in an inclined shower would have been at a smaller radius in the detector plane if the shower had been vertical.  This means that  flattening the muon LDF relative to a reference model necessarily results in a steeper CIC curve.  

The predictions for the CIC curve of two commonly used models, Q-II p and EPOS-Fe, and those of three other models, are presented in Fig.\ \ref{CSR}(lhs).  The model denoted EPOS-Mixed has a mixed 20\%p, 50\%C and 30\%Fe composition, chosen to fit the $X_{\rm max}$ distribution at 10 EeV with maximal muon signal.  It has a flatter CIC curve than EPOS-Fe, illustrating that for a given interaction model a deeper $X_{\rm max}$ distribution leads to a flatter CIC curve.  This underlines that before considering CIC predictions, one must insure that models describe the $X_{\rm max}$ distribution correctly.  Finally, comparing the EPOS-Mixed and the Boosted-$p_t$ models which have the same composition and interaction model apart from the $p_t$-boost, illustrates the argument above:  flattening the LDF to increase the muon signal results in a CIC curve with a steeper slope.   

The discussion above has shown that actually increasing the total number of muons compared to existing models is necessary to simultaneously account for both the $X_{\rm max}$ distribution and CIC curve.  How can this be done?  Defining $N_{\rm gen}$ to be the number of generations required for most pions to have energies below 100 GeV, and $f_{\rm EM}$ to be the average energy fraction in EM particles, the hadronic shower carries a fraction $f_{\rm had} \sim (1 - f_{\rm EM})^{N_{\rm gen}}$ of the total primary UHECR energy \cite{matthews05}.   Evidently, reducing either $N_{\rm gen}$ or $f_{\rm EM}$ increases $f_{\rm had}$ and hence the number of muons.   

$N_{\rm gen}$ can be reduced if, e.g., the HE interactions have higher multiplicity, hence lower secondary energies.  Then, fewer generations of hadronic interactions are required before the primary energy is degraded to the point that typical charged pions decay rather than re-interact, terminating the energy transfer to the EM component of the shower.   However this would also reduce $X_{\rm max}$ which is an independent observable that must be correctly fit.  There is no known mechanism to reduce $N_{\rm gen}$ as necessary to fit the CIC, which does not also unacceptably reduce $X_{\rm max}$.  %When $f_{\rm had}$ is decreased by changing $f_{\rm EM}$ the overall longitudinal profile is simply scaled down, causing the energy of the event to be underestimated using the standard missing-energy correction\cite{BarbosaEcal,augerCIC_ICRC11,augerEnergyCalibPLB10}, with little impact on the $X_{\rm max}$ fit.   

The second option, reducing the energy fraction carried by electromagnetically decaying particles, is non-trivial from a theoretical point of view.  Produced particle ratios for non-leading particles are to a good approximation universal in the final state of high energy processes observed in accelerator experiments, e.g., $Z^0$ decay, the central region of high energy hadron collisions, and in final states of quark-gluon-plasma probed at ALICE at the LHC.  Moreover light hadron production in conventional physics is essentially local in rapidity, so when a model is constrained to fit data at accelerator energies, $f_{\rm EM}$ does not change much with energy.   The average $f_{\pi^0}$ in $\pi^\pm$-air collisions above 10$^{16}$ eV is $\approx 0.25$ for QGSJet II and $\approx 0.2$ for EPOS 1.99 and Sibyll 2.1, with an additional $\approx 0.04$  contribution to $f_{\rm EM}$ coming from $\eta$ and $\eta^\prime$'s.   The lower $f_{\rm EM}$ of EPOS compared to QGSJet is the reason that EPOS predicts a somewhat larger muon signal than QGSJet.

\section{A toy model of the New Physics}
\label{sec:1}We now turn to identifying possible UHE phenomena with the potential to reduce the electromagnetic energy fraction in the UHECR air shower to account for observations.  Two important general observations help restrict the search.  First, while the ground muon signal varies from event to event, the variation is not dramatic, so the new phenomenon must affect almost all showers as opposed to being a rare occurrence.   This does not mean that every interaction at UHE needs to manifest new physics, since there are quite a few interactions in the VHE-UHE energy range: on average a $10^{19}$ eV primary proton shower has 2, 20 and 200 secondary collisions at energies above $10^{18}$, $10^{17}$ and $10^{16}$ eV using QGSJet II (3.3, 28 and 232 with EPOS).  Thus as long as the majority of events manifest whatever new physics is responsible for the higher muon content, and the threshold of the new physics is of order $10^{17}$ eV (14 TeV CM energy), the condition that almost all showers have increased muon content can be met.  

Second, to produce a 50-100\% increase in the muon ground signal, the new physics must impact a very substantial fraction of the total energy in UHE final states.   Production of possible new particles such as those predicted by Supersymmetry cannot account for the UHECR shower observations:  as long as they decay to a limited number of particles, their impact on the gross properties of the final states is simply too small -- even if (contrary to predictions) they were copiously produced in every interaction at UHE.   Many types of exotic particle production would actually reduce the muon signal because their decay products include missing energy which would thus be lost to the hadronic shower which yields the muons.  Production of a Quark Gluon Plasma does not substantially modify the multiplicity or particle content of final states (its most striking signature being the elliptic flow reflecting the fluid properties of the plasma) so by the analysis above, whether or not a QGP is produced in UHE proton-air collisions is largely irrelevant to the observed shower properties;  indeed QGP could be commonly produced in UHE collisions without its having a recognizable impact on shower properties.  Likewise, producing enormous $p_t$ in the first interaction, say due to production of a Black Hole with mass $\sim \sqrt{s}$, does not increase the ground muons at 1 km, unless the multiplicity is increased so much that the $X_{\rm max}$ distribution is unacceptable.

%We require a suppression of the production or decay of $\pi^0$'s.  We have not found any plausible mechanism to suppress $\pi^0$ decay, although invoking Lorentz Invariance Violation (LIV) to stabilize VHE $\pi^0$'s is technically possible.  As detailed in the Supplementary Information, the required model properties are extremely artificial.  

Chiral Symmetry Restoration (CSR) may provide a mechanism to suppress pion production, so we construct below a toy model inspired by CSR as an existence proof that a proton-only model which is fully consistent with data can be found.  In the CSR model described here, the primary UHECRs are protons, so the model also serves as a reminder of the hazards of drawing conclusions from one aspect of the showers, when other important aspects elude us -- for instance interpreting the $X_{\rm max}$ evolution with energy as implying an increasingly heavy composition, before the muon problem is resolved.

\begin{figure}[t]
\centering
\includegraphics[width=0.49\textwidth]{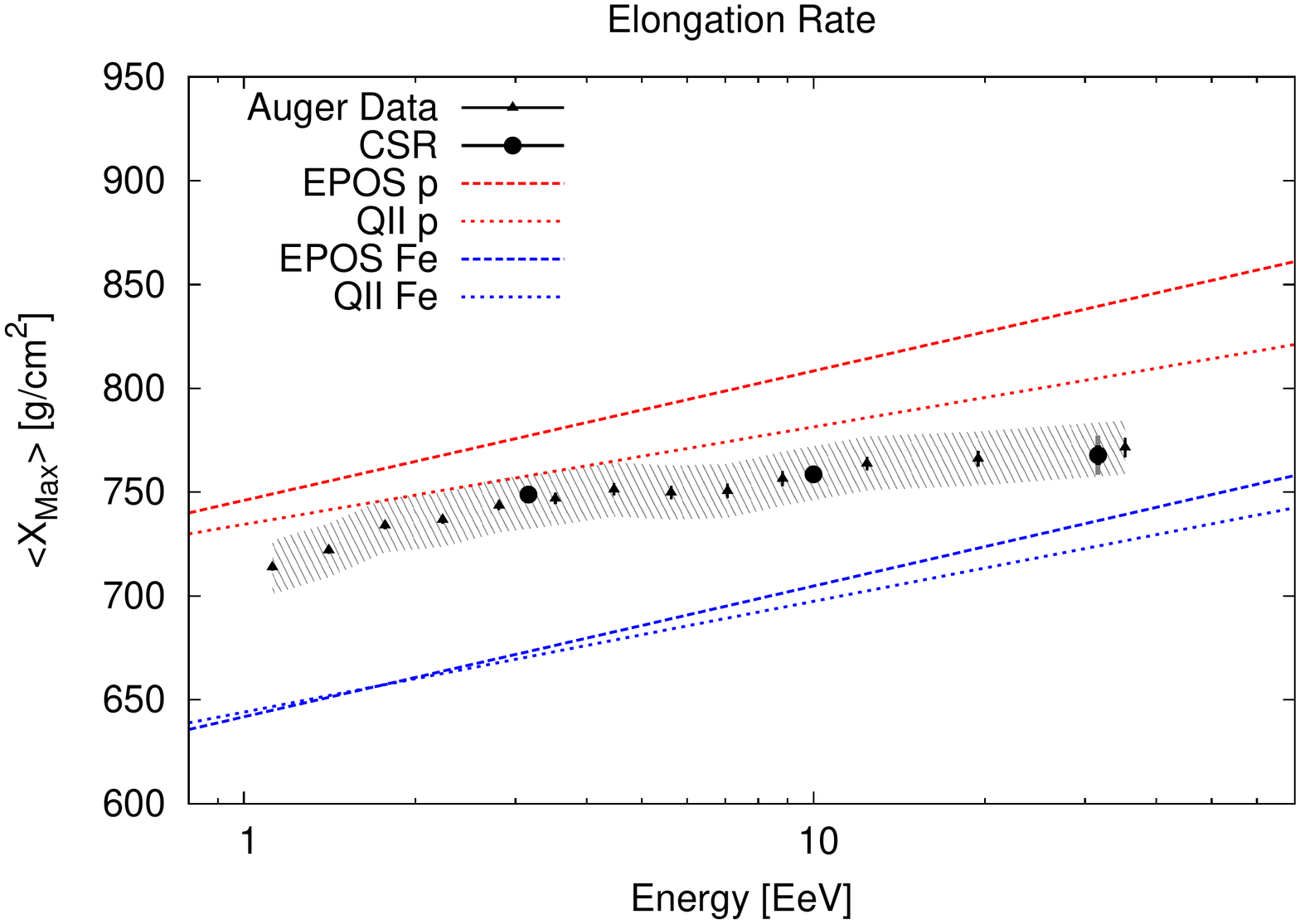}
\includegraphics[width=0.49\textwidth]{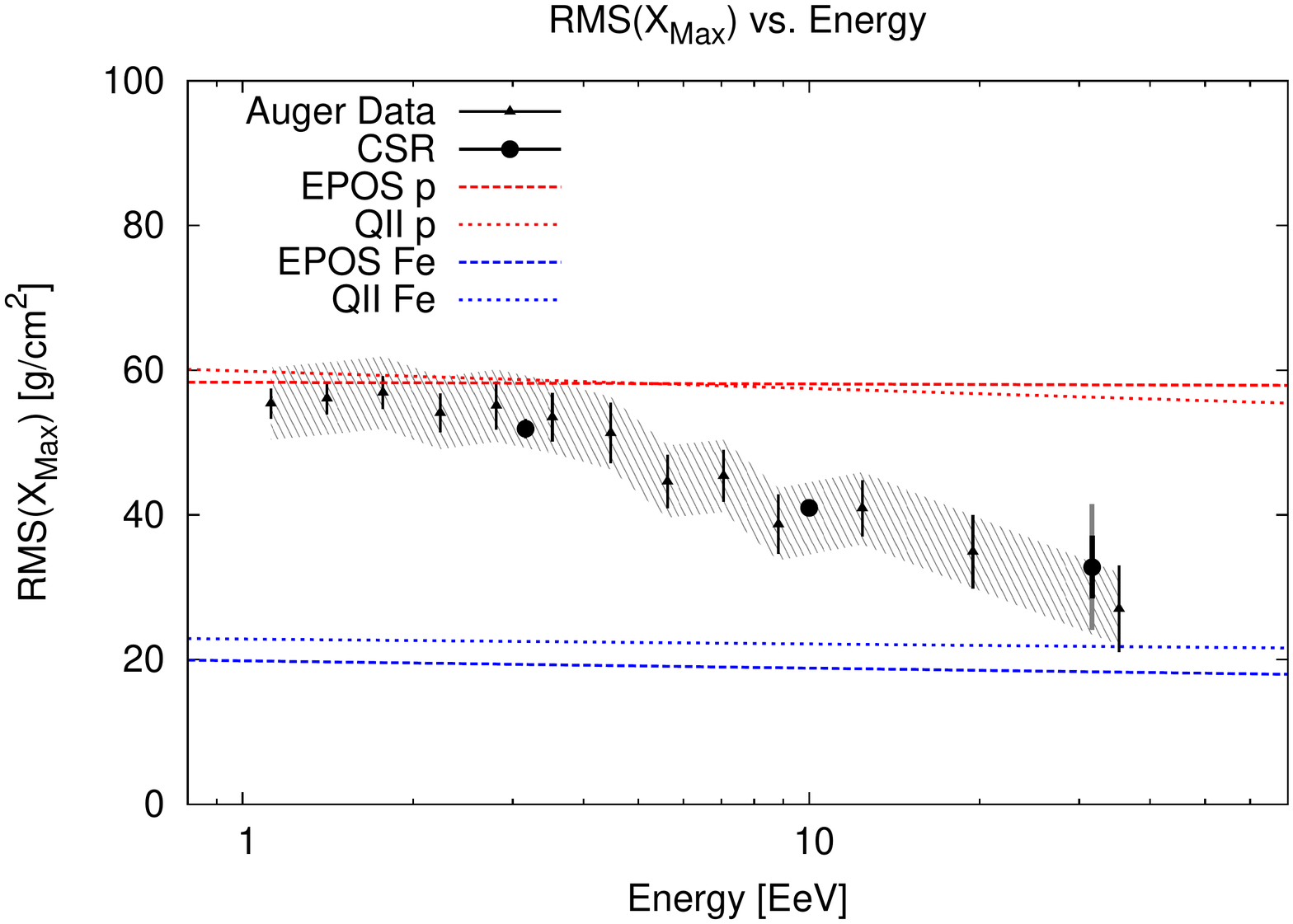}
\caption{ (lhs)$<X_{\rm max}>$ versus energy from Ref.\ \cite{augerICRCElong,augerXMaxPRL}; the systematic uncertainty in the $X_{\rm max}$ measurement is 11 ${\rm g \, cm}^{-2}$ \cite{augerXMaxPRL}.  The black (barely perceptible) and grey error bars on the highest CSR datapoint contain 68\% and 95\% of the values in different samples of 47 events.  (rhs): RMS variance in $X_{\rm max}$ as a function of energy from Ref.\ \cite{augerICRCElong,augerXMaxPRL}; the systematic uncertainty in the RMS$(X_{\rm max})$ is 5 ${\rm g \, cm}^{-2}$ \cite{augerXMaxPRL}.  The black and grey error bars on the highest CSR datapoint contain 68\% and 95\% of the values in different samples of 47 events.  \label{ElongRate}}
\vspace{+2pc}
\end{figure}

\begin{figure}[t]
\centering
\includegraphics[width=0.7\textwidth]{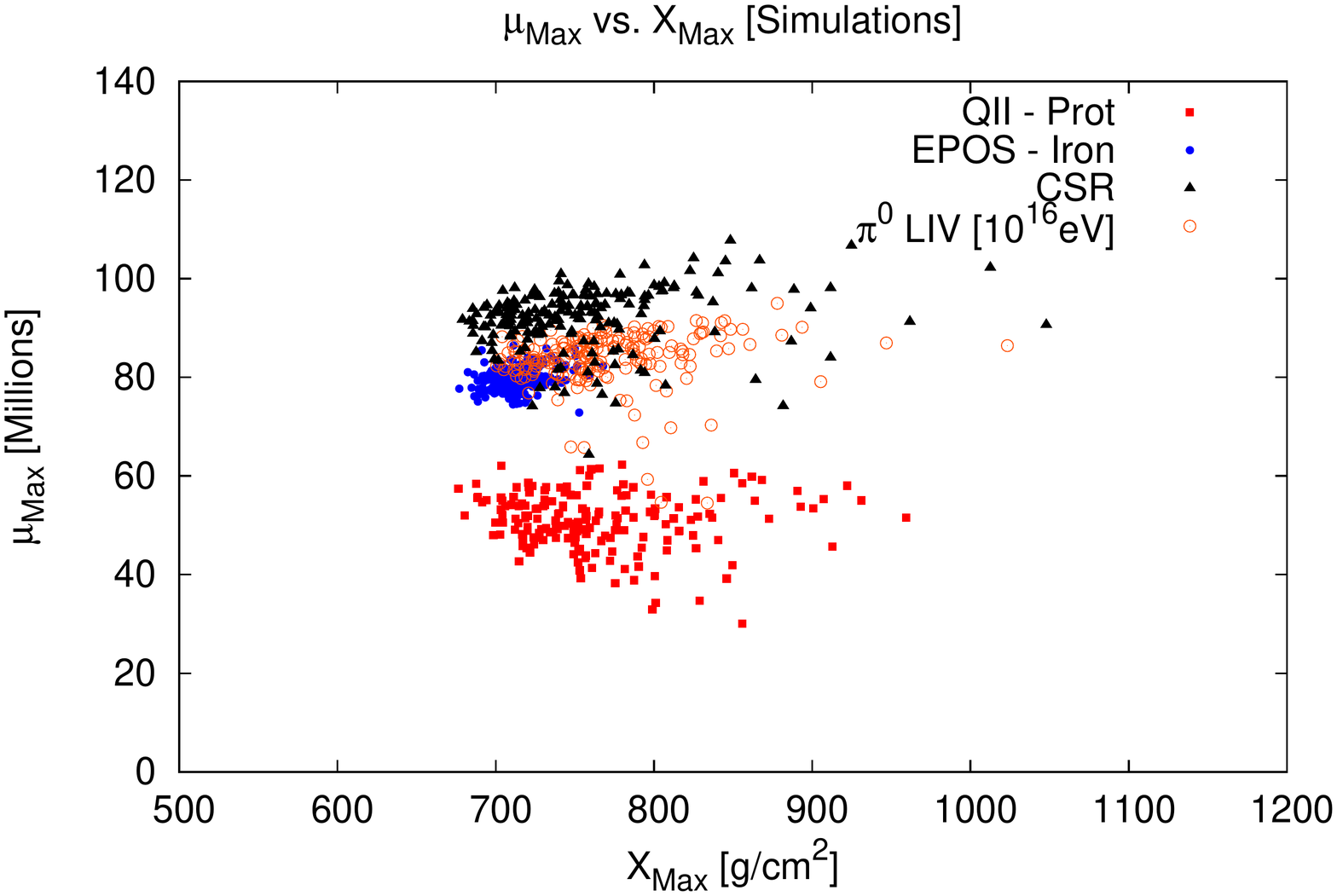}
\caption{ Scatter plot of the peak muon signal and $X_{\rm max}$ predicted for individual hybrid events, illustrating how hybrid events can be sensitive to composition and to differences between shower models through the correlation between ground signal and $X_{\rm max}$.   Simulated events for the CSR, LIV, QII-Fe and QII-p models are shown with symbols as labeled in the legend. }
\label{NmuXmax}
\vspace{+2pc}
\end{figure}

Chiral Symmetry is a near-exact symmetry of the QCD Lagrangian, because the current quark masses ($m_u < 5, \, m_d < 10$ MeV) are very small compared to the QCD scale ($\sim 100-200$ MeV).  Chiral symmetry is broken spontaneously in the normal, low temperature vacuum, due to QCD interactions which reduce the vacuum energy in the presence of a light quark--anti-quark condensate:  $<q \bar{q}>^{\frac{1}{4}} \approx 200 $ MeV.  However, at high temperature the lowest energy state has $<q \bar{q}> = 0$ and chiral symmetry is restored.   Lattice studies show a rapid cross-over of $<q \bar{q}> $ at or above the QCD deconfinement temperature, in the temperature range 160-180 MeV \cite{aokiQCDxover2006,borsanyiEOSQCD10}.

When Chiral Symmetry is spontaneously broken, Goldstone's theorem requires the existence of a light ``Goldstone boson" which is the reason for the anomalously low pion mass of $135$ MeV.  This low mass is why pions are so much more abundant than other hadrons:  their direct production is enhanced and they are the main product of the decays of hadronic resonances.  Above the Chiral Symmetry Restoration (CSR) temperature,  Chiral Symmetry can be realized either by massless fermions or parity doublets, and mesons have no reason to be light -- or even to be bound.  It should not matter whether CSR occurs via a first order phase transition or is just a rapid cross-over when the temperature of an equilibrium system is slowly increased, since UHE collisions are surely far out of equilibrium and the concept of temperature is inapplicable.  Nonetheless, we can anticipate that a shock front or bubble wall separates an inner, very high-energy-density region in which the $<q \bar{q}>$ condensate may be ``vaporized", from the exterior -- normal-vacuum -- phase.  This suggests several possible mechanisms by which CSR may suppress pion production relative to baryon--anti-baryon production, one of which is illustrated in Fig.\ \ref{CSR}(rhs).  

Theoretical studies to date have addressed static properties of the CSR phase rather than the highly non-equilibrium conditions produced in a UHE collision, so we must adopt a phenomenological approach here and develop a model of CSR in UHE collisions based on analogies with other systems.    The CSR phase is probably more readily created in central than in peripheral collisions, so we use an event generator (EPOS) to simulate a realistic distribution of peripheralities, for which we take the elasticity of the generated event to be a proxy.  When EPOS creates an event whose elasticity is above some threshold value we simply use that event as-is, while if the elasticity is below the threshold we create an event in the CSR phase as discussed below.  We took the threshold for CSR production to be $10^{17}$ eV and chose the elasticity threshold so that 50\% of the events are in the CSR phase at $10^{17}$ eV, increasing linearly in Log$_{10} E$ to 100\% at $10^{19.5}$ eV.   We expect CSR final states to be roughly similar to the final states in ordinary non-diffractive events, apart from a suppressed meson content and possibly a modified overall multiplicity, so we use EPOS to produce final states for the CSR, but convert a fraction of mesons to baryons and anti-baryons and control the average multiplicity by imposing a cap on the elasticity of CSR-phase events:  a lower elasticity cap results in a higher average multiplicity and vice versa.  Finally, since the cross-section is unconstrained except by the UHECR data we wish to fit, and some models postulate a much stronger growth with energy than EPOS, we allow for a more rapid cross-section increase than the default in EPOS.   %Fig. \ref{ModelParams} in the SI shows the values and energy dependence of the CSR model parameters used for the plots presented here.  

This first, simplistic version of a CSR model gives a remarkably good accounting of all the published shower observations discussed above.  At $10^{19}$ eV, where the statistics are good and the energy is well above the postulated energy threshold for CSR, the ground signal and $X_{\rm max}$ distributions are essentially perfectly described in every respect: the absolute magnitude and dependence on zenith angle of the ground signal (the CIC curve, Fig.\ \ref{Xmax-CIC}(rhs)) and the mean depth and shape of the peak in longitudinal development (Fig.\ref{Xmax-CIC}(lhs))\footnote{When we report predictions of the CSR model, we simulate events so they have the same calorimetric energy as in the data -- i.e., we adjust the true primary UHECR energy by the requisite factor to produce the FD signal that Auger would assign a given $E_{FD}$.}.   Furthermore, the energy dependence of the $X_{\rm max}$ distribution is also in excellent agreement with observations, as shown in Figs.\ \ref{ElongRate} (lhs) and (rhs).   
 
Our toy Chiral Symmetry Restoration model is the only model known to date which can simultaneously fit the CIC curve of ground signal versus zenith angle, the $X_{\rm max}$ distribution and its energy dependence, and the absolute calibration between SD and FD signals in hybrid events.  In the model developed here, cosmic rays above $10^{18}$ eV are pure protons and the reduction in the variance of $X_{\rm max}$ follows from the increase in the fraction of events producing the CSR phase, making the showers more homogeneous at higher energy.  The reduction in $RMS(X_{\rm max} ) $ from this effect is exhausted by $10^{19.5}$ eV, for the model parameterization used here.  If the toy CSR model is schematically correct, then either $ RMS(X_{\rm max} ) $ will plateau at higher energy, or the $p$-air cross section must increase ever more rapidly with energy than in EPOS.   The required cross-section increase is much less with even a few ${\rm g \, cm}^{-2}$ larger value of $RMS(X_{\rm max} ) $ at  $10^{19.5}$ eV, so more statistics on the $X_{\rm max}$ distribution at the highest energies will be very revealing about the CSR model.  It will also be very interesting to see, as Auger collects higher statistics, if there may be a tail of deep events due to chance occasions when an unusually large fraction of the UHE interactions are peripheral.   It should be noted that we have not yet explored the range of possible implementations of the CSR idea which are compatible with the CIC and $X_{\rm max}$ observations.  For simplicity, the toy model presented here uses a very crude scheme to decide whether a particular event will be in the normal or CSR phase, based on a hard elasticity cut, but multiplicity could be a better indicator or the impact parameter dependence of CSR production could be modeled theoretically.  Such differences in implementation may produce differences in the evolution of the $X_{\rm max}$ distribution at higher energies, or result in a different correlation between ground signal and $X_{\rm max}$, or give rise to other distinguishing characteristics which need to be identified.   

 %It must be emphasized, however, that we do not presently know whether a pure proton composition is an essential feature of the CSR model.  In a future work we will investigate whether an alternate model can be devised, possibly with different CSR parameters, to achieve an equally good fit to the ground signal and the evolution of $< X_{\rm max} > $ and $ RMS(X_{\rm max} )$, in a scenario with a mixed and changing composition.  As shown in \cite{Allardetal08}, an energy-dependent composition can be consistent with the observed spectral shape and plausible astrophysical acceleration scenarios, so an important goal is to use the evolution of the $X_{\rm max}$ distribution to infer composition evolution, if possible\cite{kampertUnger12}.  Furthermore, if there can be a mixed-composition CSR model which agrees with the showers, it would give more direct guidance as to possible LHC signatures.

The showers predicted by the CSR model differ from those of a $\pi^0$-stabilization model or of standard event generators in many details which can be tested observationally.  
A correlation is predicted between the muon content and the $X_{\rm max}$ value in individual events, which is strikingly different than for conventional event generators.  This is illustrated in Fig.\ \ref{NmuXmax} using $N_{\mu, \rm max}$, the total number of muons at maximum, as an indicator.  The CSR model likely also makes distinctive predictions for the signals in highly inclined showers and for the muon LDF, which can be tested with present data.  As experimental techniques and detector capabilities are improved, the ensemble of properties will provide discrimination between different models of UHE particle physics.   In the near future, improving statistics at higher energies and Auger's new capability to perform hybrid observations at lower energies using the SD infill and HEAT detectors, should allow the new physics and its energy dependence to be much more powerfully tested and constrained, although separating the onset of new physics from composition changes associated with the Galactic-extragalactic transition may be challenging.   The Telescope Array will provide complementary tests to Auger:  TA has both fluorescence telescopes and a surface detector array, but the TA surface detectors are scintillators rather than water tanks, hence mostly sensitive to the EM component of the showers.  Indeed TA may observe a discrepancy between the SD and FD signals in its hybrid events, relative to expectations from QGSJet-II protons \cite{TA_SDFDdiscrepancy_ICRC11} which may be explicable by the CSR model; predictions for TA observables will be presented elsewhere.

In this work we have taken an entirely phenomenological stance.  Our most important objectives are (i) to draw attention to the evidence for a new physical phenomenon of some sort, implied by the apparent change in final states in UHE particle interactions relative to accelerator energies, and (ii) to demonstrate with a concrete example, that composition evolution cannot be inferred from the energy dependence of the $X_{\rm max}$ distribution without having an accurate model of the air showers.    To do this, we have constructed a model of UHE interactions which is consistent with the key observational data.  The CSR model can be a valid phenomenological description without chiral symmetry restoration having any role in the new physical phenomena \footnote{Indeed, CSR may prove to be an acronym for something else, such as ``Completely Surprising Regime".}.  If the model has some validity, many questions need to be addressed:  What is the nature of the interface between CSR and normal phases and how does the existence of a CSR phase suppress pions?  Are only pions or all pseudoscalar mesons suppressed?  Can we predict the threshold behavior in energy and impact parameter for CSR production?  Is there a reason the CSR transition would be accompanied by an accelerating increase in cross-section?  Is there some signature of the CSR phenomenon that might be evidenced in relativistic heavy ion collisions at the LHC?  How does CSR production depend on nuclear mass?   Can particle distributions in the CSR phase be predicted from first principles?  Are there other alternatives to CSR production which are theoretically well motivated, e.g., suppression of $\pi^0$ decay by some mechanism other than Lorentz Invariance Violation?  The current generation of UHECR detectors can begin to answer some of these questions, and with the next-generation of UHECR detectors we can look forward to a dramatic increase in our understanding of fundamental physics.

%\bibliography{UHECR,CR}

\end{document}